\documentclass[preprint,10pt]{elsarticle}

\makeatletter
\def\ps@pprintTitle{%
\let\@oddhead\@empty
\let\@evenhead\@empty
\def\@oddfoot{\centerline{\thepage}}%
\let\@evenfoot\@oddfoot}
\makeatother

\journal{Chaos, Solitons \& Fractals}

\usepackage{graphicx,bbm,setspace,amssymb,amsfonts,amsmath,mathtools,mathrsfs,stmaryrd,amsthm,bm,etoolbox,comment,hyperref,url,caption,subcaption,color,enumitem,algorithm,booktabs,microtype,nicefrac,lingmacros,nccmath,tree-dvips,tikz-cd}
\usepackage[noend]{algpseudocode}
\usepackage{algorithm}

\usepackage[colorinlistoftodos]{todonotes}
\usepackage[utf8]{inputenc} 
\usepackage[T1]{fontenc}    

\DeclareMathOperator*{\argmax}{argmax}
\usepackage[noend]{algpseudocode}

\newtheorem{thm-defn}[theorem]{Theorem/Definition}

\theoremstyle{definition}

\theoremstyle{remark}

\newcommand{\ignore}[1]{}{}

\begin{document}

\begin{frontmatter}

\title{Collective infectivity of the pandemic over time and association with vaccine coverage and economic development}
   
\author[label1]{Nick James} 
\author[label2]{Max Menzies} \ead{max.menzies@alumni.harvard.edu}
\address[label1]{School of Mathematics and Statistics, University of Melbourne, Victoria, Australia}
\address[label2]{Yanqi Lake Beijing Institute of Mathematical Sciences and Applications, Beijing, China}

\begin{abstract}
This paper uses new and existing methods to study collective trends across countries throughout the pandemic, with a focus on the multivariate time series of reproduction numbers and vaccine proliferation. We begin with a time-varying analysis of the collective nature of infectivity, where we evaluate the eigenspectrum and collective magnitude of reproduction number time series on a country-by-country basis. Next, we study the topology of this eigenspectrum, measuring the deviation between all points in time, and introduce a graph-theoretic methodology to reveal a clear partition in global infectivity dynamics. Then, we compare countries' vaccine rollouts with economic indicators such as their GDP and HDI in a collective fashion. We investigate time-varying consistency and determine points in time where there is the greatest discrepancy between these indicators as a whole.  Our two primary findings are a considerable increase in collective infectivity in the latter half of the period, and a concave-up (``down then up'') pattern in the collective consistency between vaccine coverage and economic/development indicators across countries.

\end{abstract}

\begin{keyword}
COVID-19 infectivity \sep Time series analysis \sep Graph theory \sep Nonlinear dynamics \sep Rolling correlation

\end{keyword}

\end{frontmatter}

\section{Introduction}
\label{sec:Introduction}

Throughout the course of the COVID-19 pandemic, the focus of research on the virus and its infectivity has continued to shift. Initial research explored risk factors, in-patient mortality rates and recovery times \citep{Lancetdeathtime,incubation2020}, with a substantial of early analysis performed in China \citep{Jiang2020,Zu2020,Chen2020,Wang2020,Zhou2020}. Since then, medical researchers have thoroughly investigated and discovered new treatments \citep{Remdesivir,Bloch2020,toczilizumab,Cao2020}, followed by the invention of several vaccines \citep{Corey2020,Polack2020,Walsh2020}. New sophisticated methods such as deep learning and network analysis have also been recently proposed for detection and diagnosis of COVID-19 \cite{Chatterjee2023_JCS}. The aforementioned clinical research has broadened into much wider research into all the public health consequences of the pandemic, including mental health \citep{Le2020,John2020} and investigations into health care systems \citep{UShealthcare}.

Beginning slightly later than the medical research, a substantial body of research in the applied mathematics and nonlinear dynamics communities has emerged for analysing and forecasting the spread of the virus. The analytic approaches used have been broad and have continued to grow. First, many models based on classical mathematical models, such as the {reproduction number} {$R$} and the Susceptible–Infected–Recovered (SIR) model, have been proposed and systematically collated by researchers \citep{Wynants2020,ModellingEstrada2020}. These have been utilised for numerous purposes, including diagnosis and prognosis of COVID-19 patients, studies of the efficacy of medications, and vaccine development. Recent work has specifically focused on estimating reproduction numbers in the context of vaccination programs, and investigating their impact \citep{Xavier2022_JCS}.

Next, nonlinear dynamics researchers have proposed many sophisticated extensions to the classical predictive SIR model, including analytic techniques to find explicit solutions \citep{SIRBarlow2020, SIRWeinstein2020}, modifications to the model with additional variables \citep{SIRNg2020,SIRVyasarayani2020,SIRCadoni2020,SIRNeves2020,SIRComunian2020,Sun2020}, incorporation of Hamiltonian dynamics \citep{SIRBallesteros2020} or network models \citep{SIRLiu2021}. Some researchers have even modified SIR models to examine not just COVID-19 counts themselves but economic and other follow-on impacts of the pandemic \citep{Georgiev2023_JCS}. In more recent times, numerous SIR models have been developed to incorporate the dynamics of vaccination \cite{Ahumada2023_CSF,Meng2023_CSF,Paul2023_CSF,Angeli2022_CSF}, including network models \cite{Chatterjee2023_CSF}.

Other mathematical approaches to analysis and prediction include power-law models \citep{Manchein2020,Blasius2020,Beare2020,brugnago_how_2020}, forecasting models \citep{Perc2020,Ribeiro2020_Brazil1,Ribeiro2020_Brazil2}, fractal approaches \citep{Boccaletti2020,Castillo2020,Castillo2021}, neural networks \citep{Melin2020,Ali2023_CSF}, stochastic approaches \cite{Xu2023_CSF,Olivares2023_CSF}, methods from Bayesian statistics \citep{james2021_spectral,Manevski2020},  distance analysis \citep{James2021_virulence}, network models \citep{Shang2020, Karaivanov2020,Hncean2020_Romania,Li2021_Matjaz,Gosak2021,Markovi2021}, analyses of the dynamics of transmission and contact \citep{Saldaa2020,Danchin2021}, clustering \citep{James2020_chaos,Machado2020} and many others \citep{james2021_TVO,Nraigh2020,Glass2020}. This community of researchers is inherently interdisciplinary, so have frequently studied other impacts of COVID-19 outside epidemiology, such as its effect on financial and cryptocurrency markets \citep{Akhtaruzzaman2020,Akhtaruzzaman2020_2,Jamesfincovid,Zaremba2020,Okorie2020,James2021_crypto,Lahmiri2020,James2021_crypto2,Mnif2020,james2021_mobility,Goodell2020,arouxet2020}. Finally, numerous more recent articles have been devoted to understanding the spatial components of the virus' spread, in numerous countries \citep{Zhou2020_covidUS,Melin2020_2,Wang2023_CSF,James2021_geodesicWasserstein,Wang2020_spatioUS,cuellar_excess_2022,Shi2023_CSF}.

In more recent times, the pandemic and its associated research have taken a different form. Most developed countries have essentially completed their vaccine rollout, and yet cases continue to exhibit new peaks throughout many different regions \citep{james2021_CovidIndia}. In addition, most countries have reduced their restrictions and opened their borders, meaning they are no longer isolated ecosystems. Thus, a new body of work has emerged with an emphasis on necessary collaboration between countries to mitigate ongoing risks of COVID-19 \citep{Priesemann2021_actionplan,Priesemann2021_Eurostrategy,Priesemann2021}. Vaccines remain an integral part of that response, so ongoing research is necessary to determine their impact on cases and effectiveness \citep{Piraveenan2021}. On this question, there  is a substantial amount of medical research \citep{spensley_comparison_2022,tregoning_progress_2021,lisewski_effectiveness_2021} and more recently some mathematical and nonlinear dynamics works examining trends \citep{Sharma2021}. Of particular note are works that investigate the association between vaccination and infectivity, either in a single country \cite{Meng2023_CSF,Paul2023_CSF} or individually across several countries \cite{Wang2023_CSF,Angeli2022_CSF}.

Our work is heavily inspired by the aforementioned trends in the recent literature, but we take a different approach. Existing work has discussed and called for further cooperation between countries to continue mitigating COVID-19 after the vaccine rollout, and some work referenced above has analysed infectivity on the basis of one country at a time. However, we are unaware of any mathematical work has examined collective trends in infectivity and relationships between countries on a worldwide basis, especially incorporating the interplay with the vaccine rollout. Specifically, we are unaware of any research within the framework of time series analysis that studies collective behaviour of infectivity time series on a country-by-country basis.

Our paper is structured as follows. Section \ref{sec:data} summarises all the data we draw upon in this work. In Section \ref{sec:Time_varying_infectiousness}, we study collective dynamics of various countries' reproduction rate time series. We use methods of time series analysis typically used in financial and other applications \citep{Gopikrishnan2001,james_arjun,Laloux1999,james2021_MJW,Plerou2002}. In particular, we notice an abrupt increase in collective dynamics from approximately March/April 2021. In Section \ref{sec:Graph_theoretic_partition}, we analyse this more closely with graph-theory-inspired methods, identifying a point in time that exhibits maximal separation in the eigenspectrum. In Section \ref{sec:Vaccine_rollout_consistency}, we study the time-varying {collective} consistency between countries' vaccine proliferation and their economic and human development. {Section \ref{sec:Conclusion} summarises our findings regarding collective behaviours across countries throughout the pandemic and discusses possible interpretations and limitations thereof.}

\section{Data}
\label{sec:data}

All the data used in this paper are obtained from Our World in Data (\url{https://ourworldindata.org}). We select the $N=50$ countries with the greatest total reported case counts as of June 2022 and consider  {two multivariate} time series:  {reproduction numbers}  {$R_t$}, according to their method of estimation \citep{arroyo-marioli_tracking_2021}, and counts of fully vaccinated individuals. Our data spans a period of 1 April 2020 to 1 May 2022, in total $T=761$ days. We end our data at this point: beyond  here (the northern hemisphere summer of 2022), numerous countries stop their regular collection and reporting of data, shifting in succession to weekly rather than daily case counts \cite{CDC_Covid_ending,UK_Covid_ending,Aus_Covid_ending}. {We also use countries' human development indices (HDI) and gross domestic product (GDP) drawn from the same data source.}

{We briefly clarify and disambiguate the definition and selection of the reproductive number time series $R_t$, as several definitions exist. First, the \emph{basic reproduction number} $R_0$ is a fixed real number that estimates the average number of cases caused by a primary case when the population is fully susceptible. The \emph{effective reproduction number} $R_t$ is a time-varying quantity that tracks the average number of secondary cases infected by each primary case as an epidemic continues. However, there are two slightly different precise definitions of the effective reproduction number: the \emph{instantaneous reproduction number} or the \emph{case reproduction number} - ``the instantaneous reproductive number measures transmission at a specific point in time, whereas the case reproductive number measures transmission by a specific cohort of individuals'' \citep{Gostic2020}.}

{\cite{Fraser2007} explains, ``The case reproduction number $R_c(t)$ is a property of individuals infected at time $t$, and is the average number of people someone infected at time $t$ can expect to infect'' while the instantaneous number is ``the average number of people someone infected at time $t$ could expect to infect should conditions remain unchanged''. Based on this distinction as well as Eq. (8) and his subsequent remark, ``the case reproduction number is a smoothed function of the instantaneous reproduction number,'' we believe the case reproduction number is smoother and more suitable for analysis. It is this quantity that our data source uses, adapting the Kalman smoother proposed by \cite{arroyo-marioli_tracking_2021}. {To be precise, the reproduction number time series data hosted on Our World in Data are precisely the case reproduction number time series $R_c(t)$  calculated by \cite{arroyo-marioli_tracking_2021}; we draw this data without any further modification for all analysis in this paper.} Thus, throughout this manuscript, our time series $R_t$ use the case reproduction number, one of two specific forms of the more frequently used terminology that is the effective reproduction number.
}

\section{Time-varying collective infectivity}
\label{sec:Time_varying_infectiousness}

Let $R_i(t)$ be the multivariate time series of daily {reproduction number} ({$R_t$}) values for countries listed alphabetically and indexed $i=1,...,N, t=1,...,T$. We begin by choosing a window of $\tau=90$ days over which we compute correlations between rolling {$R_t$} time series. This approach is inspired by analyses of correlations and distances between financial time series \citep{james2021_portfolio,Fenn2011,Wilcox2007,Pan2007,james2022_stagflation}, which demonstrate that stock market
efficiency is a collective phenomenon \cite{Kim2005,james_georg,Alves2020,James2023_cryptoGeorg,Heckens2020,financialcrises} These are defined as follows:  for $t=\tau,...,T$, define an $N \times N$ matrix $\bm \Psi(t)$ by
\begin{align}
\label{eq:rhodefn}
    \Psi_{ij}(t)=\frac{1}{\tau} \frac{\sum_{s=t-\tau+1}^t (R_i(s) - \langle R_i \rangle)(R_j(s) - \langle{R}_j \rangle)}{\left(\sum_{s=t-\tau+1}^t (R_i(s) - \langle R_i \rangle)^2 \right)^{1/2} \left( \sum_{s=t-\tau+1}^t (R_j(s) - \langle R_j \rangle)^2\right)^{1/2}}, \\ i,j=1,...,N,
\end{align}
where $\langle . \rangle $ denotes the temporal average over the time window $[t-\tau+1,t]$. We can gain additional understanding of this matrix by decomposing it as follows:
\begin{align}
\label{eq:alternativerho}
    \bm \Psi(t)= \frac{1}{\tau} {\tilde{R}} {\tilde{R}}^T.
\end{align}
In the equation (\ref{eq:alternativerho}) above, $\tilde{R}$ is a $N \times \tau$ matrix of standardised {$R_t$} values obtained by rescaling by the mean $\langle . \rangle $ and standard deviation $\sigma(.)$ over the interval $[t-\tau+1,t]$, namely $\tilde{R}_i(s) = [R_i(s) - \langle R_i \rangle] / \sigma(R_i)$.

With this matrix decomposition, we may deduce that $\bm \Psi(t)$ is a symmetric positive semi-definite matrix whose entries lie in $[-1,1]$. It carries real and non-negative eigenvalues $\lambda_i(t)$ that we can reorder by $\lambda_1 \geq \lambda_2 \geq ... \geq \lambda_N$. These convey the magnitude of different axes of variation within $\bm \Psi(t)$. As all diagonal entries of $\bm\Psi$ coincide with 1, the trace of $\bm\Psi$ is equal to $N$. Thus, we may normalise the eigenvalues by defining $\tilde{\lambda}_i = \frac{\lambda_i}{\sum^{N}_{j=1} \lambda_j}= \frac{\lambda_i}{N}$. In particular, $\tilde{\lambda}_1(t)$ coincides with a normalised \emph{operator norm} \citep{RudinFA}:
\begin{align}
\label{eq:lambda1}
    \tilde{\lambda}_1(t) = \frac{1}{N} \| \bm \Psi(t) \|_{op} = \frac{1}{N} \sup_{v \in \mathbb{R}^N - \{0\}} \frac{\|\Psi(t)v \|}{\|v\|}.
\end{align}
{Like other \emph{matrix norms}, this operator norm is a scalar quantity that reflects the total collective size of the matrix coefficients. For example, $\| \lambda A \|_{op} = \mid \lambda \mid \|  A \|_{op}$  for any matrix $A$ and scalar $\lambda$. In our context, $\tilde{\lambda}_1(t)$ gives a measure of the overall size of the matrix $\bm \Psi(t)$ or the collective strength of correlations. That is, a larger value of $\tilde{\lambda}_1(t)$ reflects broadly larger values of the matrix $\bm \Psi(t)$, and hence broadly larger collective correlations between reproduction number time series. So $\tilde{\lambda}_1(t)$ is a time-varying scalar function that summarises how similarly (in a collective sense) are the behaviours of reproduction number time series across our collection of countries.}

\begin{figure}
    \centering
    \includegraphics[width=\textwidth]{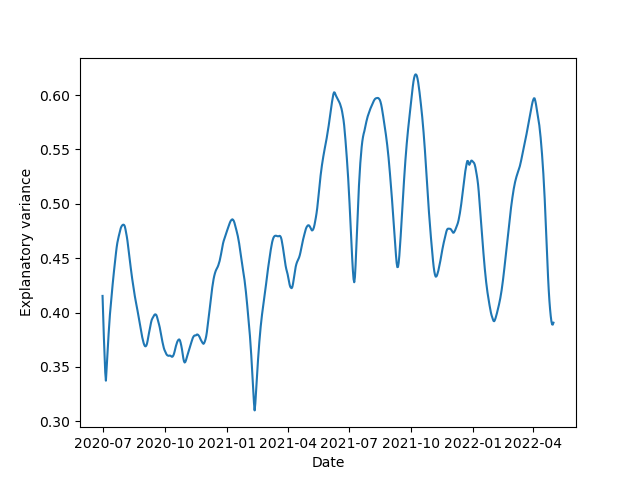}
    \caption{Trends in the normalised first eigenvalue  $\tilde{\lambda_1}(t)$, as defined and discussed in (\ref{eq:lambda1}). Each indexed date provides a measure of the collective strength of correlations between  {reproduction number} time series over the prior 90 days. We observe a sharp translation upward during the latter half of the period analysed. In addition, there is the emergence of wave-type behaviours in the strength of collective infectivity.}
    \label{fig:Lambda_1_t}
\end{figure}

\begin{figure}
    \centering
    \includegraphics[width=\textwidth]{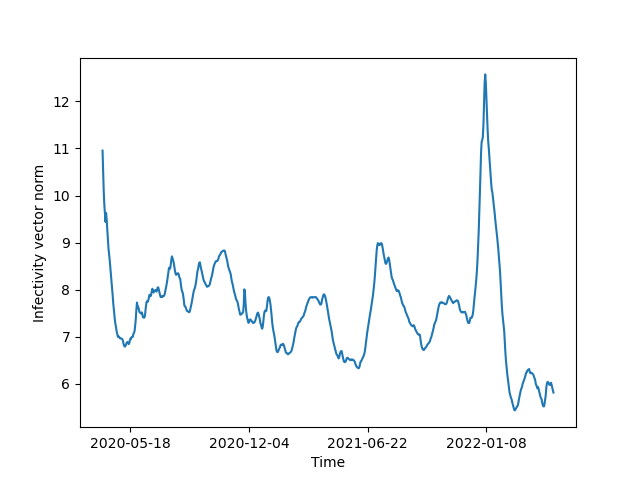}
    \caption{Trends in the Frobenius vector norm $\nu_R(t)$, as defined in (\ref{eq:R0vector}). Each indexed date provides a measure of the collective magnitude of {reproduction numbers} on a particular day. We observe peaks in the first wave of COVID-19 across the world in April 2020, and the global dominance of the omicron in early 2022.}
    \label{fig:R0_vectornorm}
\end{figure}

We display the function $\tilde{\lambda}_1(t)$ in Figure \ref{fig:Lambda_1_t}. This shows the evolution of the collective strength of correlations between {reproduction number} time series during our window of analysis. The shape of the function tells an interesting story, with three key insights. First, there is significant variability in the range of collective behaviour of countries' infectivity, with the value of $\tilde{\lambda}_1(t)$ ranging from $\sim 0.3$ to $\sim 0.6$. Second, there is clearly a sharp increase (translation upward) in global collective behaviours around March-April 2021. It is likely that this period corresponds to two key themes. This period is broadly consistent with the onset of the delta variant, which was significantly more infectious that prior strains of COVID-19. Furthermore, many countries began lifting restrictions, with a larger proportion of their populations having been vaccinated. This likely led to a greater number of susceptible candidates transmitting COVID-19 to one another in various countries and hence greater collective uniformity and correlation strength between countries. The final notable insight is the significant increase in the oscillatory behaviour of the function beyond March-April 2021. There are at least 5 ``waves'' of collective infectivity in the latter part of our graph. This may be indicative of the ``stop-start'' nature of many countries' economic policies with regards to their gradual reopening, and the gradual evolution of the COVID-19 virus.

We complement the above analysis, with focuses on collective correlations between {reproduction numbers}, with an examination of the collective magnitude across countries' {$R_t$} time series. Specifically, let 
\begin{align}
\label{eq:R0vector}
\nu_R(t)= \left(\sum_{i=1}^N R_i(t)^2 \right)^\frac12, t=1,...,T
\end{align}
be the Euclidean norm of the vector of  {reproduction numbers} $(R_1(t),...,R_N(t)) \in \mathbb{R}^N$ at each time. Larger values of $\nu_R(t)$ reflect greater collective magnitudes of infectivity across the world (while $\tilde{\lambda}_1(t)$ pertains to greater collective correlations). We note that $\nu_R(t)$ can be computed at any time $t=1,...,T$, whereas $\tilde{\lambda}_1(t)$ is computed over rolling intervals indexed $t=\tau,...,T$.

The function $\nu_R(t)$ is displayed in Figure \ref{fig:R0_vectornorm}. Again, several insights can be gleaned. Countries' collective magnitude of their infectivity time series were largest at two key times: early 2020, when the first wave of COVID-19 cases exploded across the world, severely affecting for the first time countries other than China and Italy \citep{al_jazeera_2020}; and early 2022, when the omicron variant resulted in high  {reproduction numbers} across the world. Intriguingly, the marked shift in collective correlation dynamics in March/April 2021 precedes the omicron variant, neither does omicron appear to have affected collective correlations significantly. It did, of course, affect collective magnitudes in  {reproduction numbers}.

\section{Graph-theoretic partition}
\label{sec:Graph_theoretic_partition}

In the above section, we highlight the marked change in collective homogeneity of infectivity (summarised by $\tilde{\lambda}_1(t)$ of rolling {$R_t$} time series) around March/April 2021. In this section, we aim to elucidate and quantify this notable change point via a novel approach informed by graph theory. We further analyse the previous section's primary object of study, the time series $\tilde{\lambda}_1(t), t=\tau,...,T$. First, for $s,t=\tau,...,T$, we construct a distance that distinguishes between such values, $d(s,t)= \mid \tilde{\lambda}_1(s) - \tilde{\lambda}_1(t) \mid $. This produces a $(T-\tau+1) \times (T-\tau+1)$ distance matrix $D_{st}=d(s,t)$. For notational convenience, let $T_1=T-\tau+1.$ We can also associate a corresponding $T_1 \times T_1$ \emph{affinity matrix} defined by
\begin{align}
    \label{eq:affinity}
    A_{st}=1 - \frac{D_{st}}{\max D}, s,t=\tau,...,T.
\end{align}
We define the primary change point in the time series $\tilde{\lambda}_1(t)$ as follows:
\begin{align}
    \label{eq:primarycp1}
    T_0 = \argmax_p f(p); \\
    f(p)= \frac{1}{p(T_1 - p)} \sum_{s\leq p, t>p} D_{st}.
    \label{eq:f_defn}
\end{align}
That is, $T_0$ is chosen to maximise the normalised distance (calculated from $D$) between times before and after $T_0$.

This approach is inspired by graph theory and can be interpreted as follows. Let $G=(V,E)$ be a (weighted) graph with vertex set $V=\{\tau,...,T\}$ and edge set $E$ where every two $s,t$ are connected by an edge of weight $d(s,t)$. Then $T_0$ is chosen subject to two constraints:
\begin{itemize}
    \item We wish to partition the vertex set into $V = V_1 \cup V_2$ such that $V_1$ and $V_2$ are disjoint and unbroken time intervals;
    \item subject to the above, we wish to find a \emph{maximum cut} of the graph subject to its edge weighting.
\end{itemize}

For the sake of robustness, we also present an alternative method to determine a primary change point. As an alternative approach, we define
\begin{align}
    \label{eq:primarycp2}
    S_0 = \argmax_p g(p); \\
    g(p) = \frac{1}{p^2} \sum_{s, t\leq p} A_{st} + \frac{1}{(T_1 - p)^2} \sum_{s, t> p} A_{st}.
    \label{eq:g_defn}
\end{align}
Here $S_0$ is chosen to maximise the (normalised) collective affinity between times both before and after $S_0$. This definition is also inspired by graph theory, but aims to find a \emph{minimum cut} of the graph subject to an alternative edge weighting where every two $s,t$ are connected by an edge of weight $A_{st}$, their affinity.

\begin{figure}
    \centering
\begin{subfigure}[b]{\textwidth}
    \includegraphics[width=\textwidth]{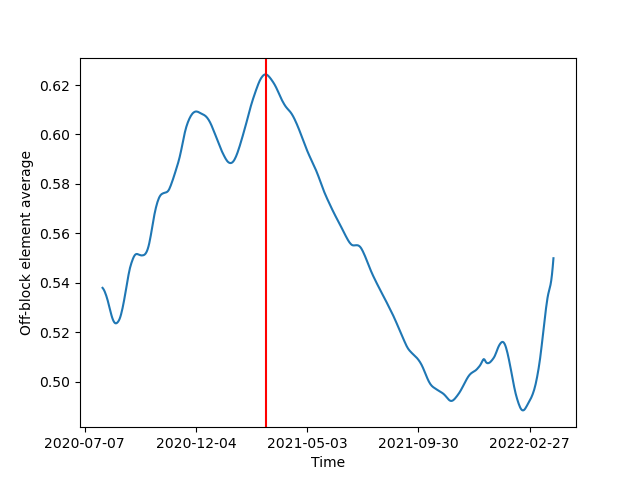}
    \caption{}
    \label{fig:Max_flow_min_cut1}
\end{subfigure}
\begin{subfigure}[b]{\textwidth}
    \includegraphics[width=\textwidth]{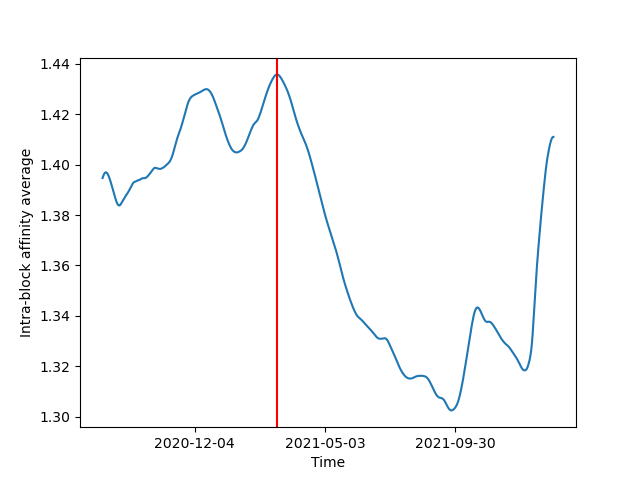}
    \caption{}
    \label{fig:Max_flow_min_cut2}
\end{subfigure}
\caption{Trends in the functions (a) $f(p)$, defined in (\ref{eq:f_defn}), and (b) $g(p)$, defined in (\ref{eq:g_defn}). These measure the normalised distance between and affinity among the values of the first eigenvalue corresponding to times before and after time $p$, respectively. Vertical red lines denote the respective maximal values, defined as $T_0$ and $S_0$ respectively, as in Section \ref{sec:Graph_theoretic_partition}. The similarity of the observed maxima demonstrates the robustness of our methodologies to identify the cleanest separation in the evolution of $\tilde{\lambda}_1(t).$}
\label{fig:Max_flow_min_cut}
\end{figure}

In Figure \ref{fig:Max_flow_min_cut}, we plot both the aforementioned functions $f(p)$ and $g(p)$ as well as their respective maxima, $T_0$ and $S_0$, as defined above. The considerable similarity in the figures demonstrates the robustness of the method, and together they reveal the point in time corresponding to the most abrupt change in collective infectivity. The red vertical lines correspond to the points $T_0$ and $S_0$ in time that maximise the aforementioned separation in our graph's structure. This maximum corresponds to approximately March/April 2021, and reflects the sharp increase in collective behaviours during the second half of our period of analysis. 

\begin{figure}
    \centering
    \includegraphics[width=\textwidth]{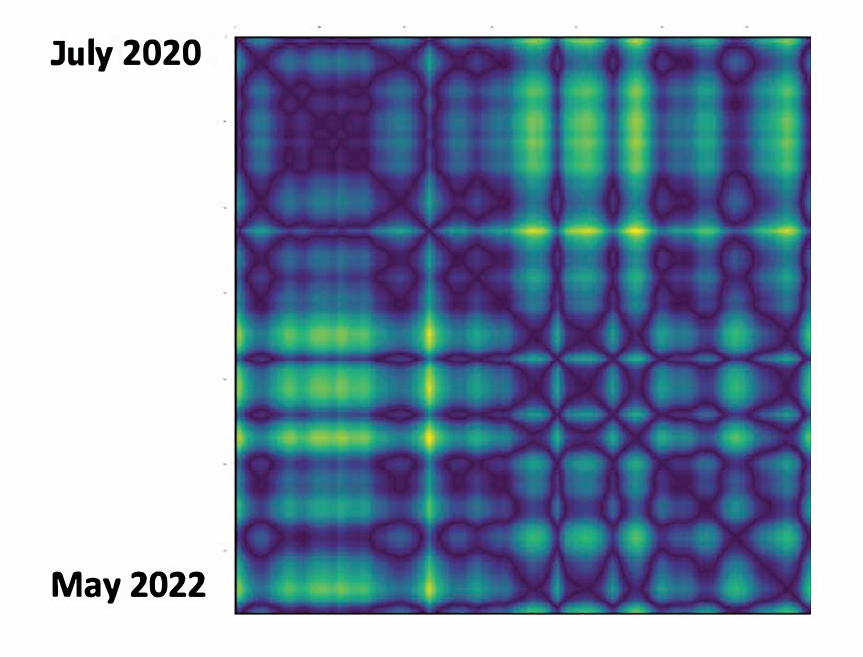}
    \caption{Heat map depicting values of $d(s,t)$ determining the distance between the normalised first eigenvalue $\tilde{\lambda}_1(t)$ at all points $t=\tau,...,T.$ As each value of $\tilde{\lambda}_1(t)$ reflects behaviour over the previous 90 days, the figure begins with July 2020. Lighter values indicate greater values of $d(s,t)$.}
    \label{fig:Distance_1}
\end{figure}

\begin{figure}
    \centering
    \includegraphics[width=\textwidth]{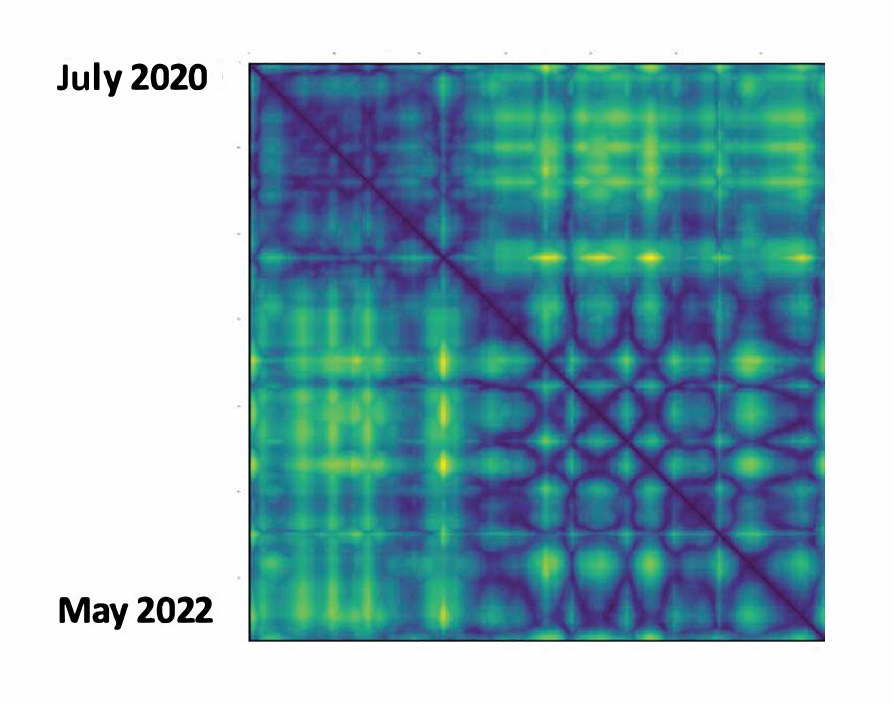}
    \caption{Heat map depicting values of $d^{10}(s,t)$ determining the distance between the normalised first ten eigenvalues $\tilde{\lambda}_i(t), i=1,...,10$ at all points $t=\tau,...,T.$ As each value of $\tilde{\lambda}_i(t)$ reflects behaviour over the previous 90 days, the figure begins with July 2020. Lighter values indicate greater values of $d^{10}(s,t)$.}
    \label{fig:Distance_10}
\end{figure}

Figure \ref{fig:Distance_1} provides more visual detail regarding the distance $d(s,t)=\mid \tilde{\lambda}_1(s) - \tilde{\lambda}_1(t) \mid$. This figure displays all values of the distance for $s,t=\tau,...,T.$ Such a heat map can reveal clear patterns in the distance between elements and offer the possibility of applying methods such as hierarchical clustering \citep{Ward1963,Szekely2005,James2020_nsm,Mllner2013}. We also plot an alternative distance in Figure \ref{fig:Distance_10}, defined by $d^{10}(s,t)=\sum_{k=1}^{10} \mid \tilde{\lambda}_l(s) - \tilde{\lambda}_k(t) \mid$, incorporating the first ten normalised eigenvalues. Figures \ref{fig:Distance_1} and \ref{fig:Distance_10} convey a similar story, with two sub-matrices exhibiting high self-similarity. Both figures are consistent with our finding in Figure \ref{fig:Max_flow_min_cut}, and suggest a clear break in the structure of the $d(s,t)$ matrix corresponding to March/April 2021. However, there are several subtle differences between Figure \ref{fig:Distance_1} and Figure \ref{fig:Distance_10} that are worth highlighting. First, Figure \ref{fig:Distance_1}, which displays distances between $\tilde{\lambda_1}$ at various points in time, consists of a sharper distinction between the two sub-matrices. By contrast, Figure \ref{fig:Distance_10}, which computes distances between the first ten normalised elements of the eigenspectrum $\tilde{\lambda}_1,...,\tilde{\lambda}_{10}$, displays a more diffuse transition between the two periods. The second key insight, which is highly consistent between Figures \ref{fig:Distance_1} and \ref{fig:Distance_10}, is the interesting geometry displayed in the secondary sub-matrix. This feature reflects the periodic behaviour of the eigenspectrum following the primary change point in March/April 2021.

\section{Vaccine rollout consistency}
\label{sec:Vaccine_rollout_consistency}
\begin{figure*}
    \centering
    \begin{subfigure}[b]{0.49\textwidth}
        \includegraphics[width=\textwidth]{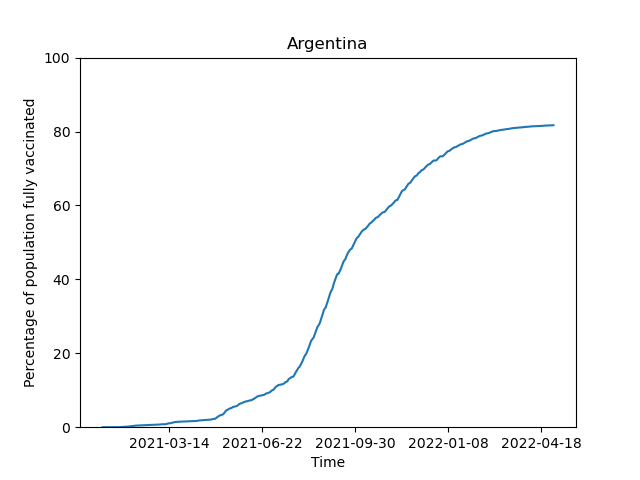}
        \caption{}
        \label{fig:Vaccination_argentina}
    \end{subfigure}
    \begin{subfigure}[b]{0.49\textwidth}
        \includegraphics[width=\textwidth]{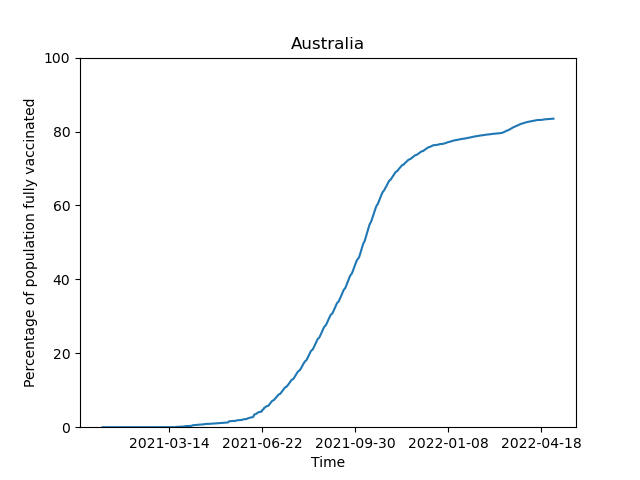}
        \caption{}
        \label{fig:Vaccination_australia}
    \end{subfigure}
    \begin{subfigure}[b]{0.49\textwidth}
        \includegraphics[width=\textwidth]{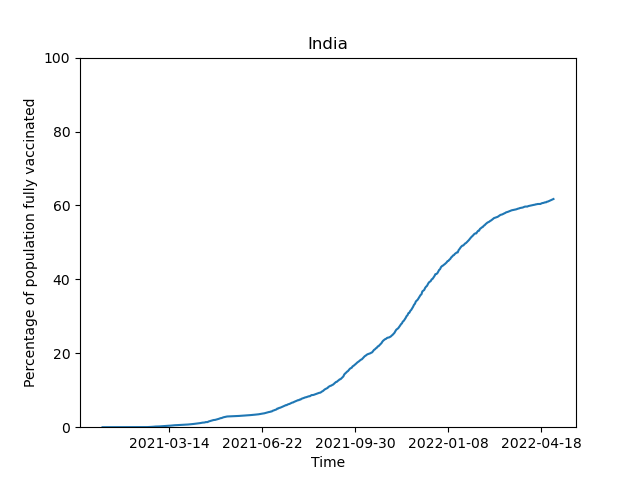}
        \caption{}
        \label{fig:Vaccination_India}
    \end{subfigure}
    \begin{subfigure}[b]{0.49\textwidth}
        \includegraphics[width=\textwidth]{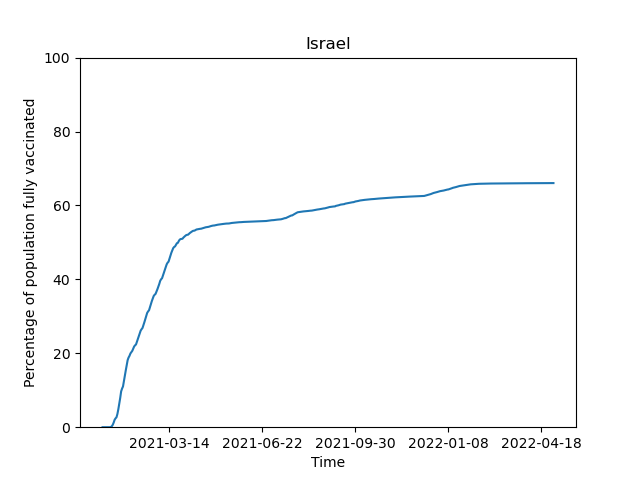}
        \caption{}
        \label{fig:Vaccination_israel}
    \end{subfigure}
    \begin{subfigure}[b]{0.49\textwidth}
        \includegraphics[width=\textwidth]{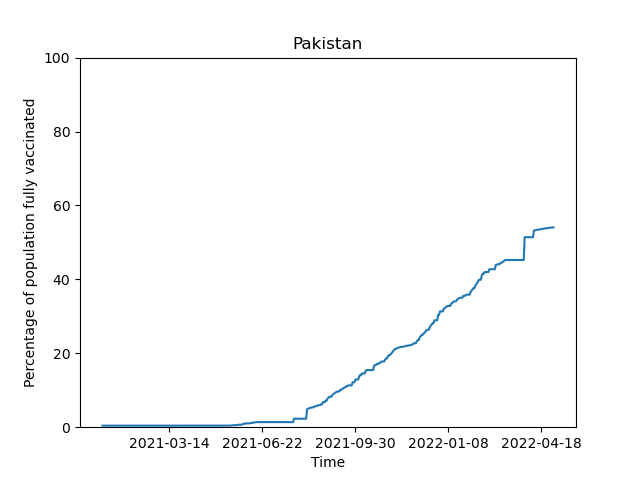}
        \caption{}
        \label{fig:Vaccination_pakistan}
    \end{subfigure}
    \begin{subfigure}[b]{0.49\textwidth}
        \includegraphics[width=\textwidth]{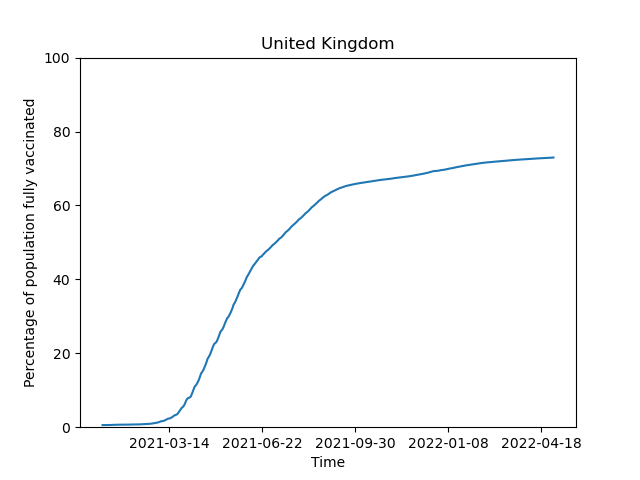}
        \caption{}
        \label{fig:Vaccination_UK}
    \end{subfigure}
    \caption{Time-varying rate of the population that is fully vaccinated for (a) Argentina, (b) Australia, (c) India, (d) Israel, (e) Pakistan and (f) the United Kingdom. Countries that were more efficient in fully vaccinating their population, such as Israel, have a steeper gradient, at an earlier point in time. Countries that were less efficient, such as Pakistan, have a less steep gradient commencing at a later point in our analysis window.}
    \label{fig:Vaccination_rates}
\end{figure*}

In this section, we study the collective consistency between countries' vaccine rollouts and economic indicators such as their gross domestic product (GDP) and human development index (HDI). To do so, we introduce the following variables. Let $v_i(t)$ be the multivariate time series that records the fully vaccinated percentage of a country's population over time,  $i=1,...,N$ and $t=1,...,T$. Our country-specific GDP and HDI scores are indexed $g_i$ and $h_i$ respectively, $i=1,...,N$. As we wish to restrict attention to the period of global vaccine proliferation, this section restricts analysis from 1 Febuary 2021 to 1 May 2022, a period of $T=455$ days. For purposes of illustration, we plot the trajectory of $v_i(t)$ for select countries in Figure \ref{fig:Vaccination_rates}. To explore the time-varying consistency between countries' vaccination rollout and their GDP and HDI, we proceed as follows. At each time $t$, we construct the following distance matrices:
\begin{align}
    D^{GDP}_{ij} = \mid g_i - g_j \mid; \\
    D^{HDI}_{ij} = \mid h_i - h_j \mid; \\
    D(t)^{V}_{ij} = \mid v_i(t) - v_j(t)  \mid.
\end{align}
We then convert each distance matrix into an \emph{affinity matrix} using the same definition as (\ref{eq:affinity}):
\begin{align}
    A^{GDP}_{ij} = 1 - \frac{D_{ij}^{GDP}}{\max\{ D^{GDP} \}}; \\
    A^{HDI}_{ij} = 1 - \frac{D_{ij}^{HDI}}{\max\{ D^{HDI} \}}; \\
    A^{V}_{ij}(t) = 1 - \frac{D_{ij}^{V}(t)}{\max\{ D^{V} (t) \}}, i,j=1,...,N,t=1,...,T. 
\end{align}
These affinity matrices are appropriately normalised and can be compared directly to study the consistency between vaccine proliferation and HDI/GDP. We generate two \emph{inconsistency matrices} as follows:
\begin{align}
    \text{INC}^{HDI,V}(t) =A^{V}(t) - A^{HDI}; \\
    \text{INC}^{GDP,V}(t) =A^{V}(t) - A^{GDP}.
\end{align}
A larger absolute value of an entry in the matrix $\text{INC}^{HDI,V}(t)$ indicates that the relationships between two countries regarding HDI and their vaccination percentage at a point in time are quite different, analogously for $ \text{INC}^{GDP,V}(t)$. To explore the collective consistency between such attributes across our entire collection, we calculate the $L^1$ norm of our resulting consistency matrices and study how they evolve over time. For a $N \times N$ matrix $A$, its $L^1$ norm is defined by  $\|A\|=\sum_{i,j=1}^N  \mid A_{ij} \mid  $. We compute the following {two} matrix norms, {which we term collective inconsistency functions (functions of time)}:
\begin{align}
\label{eq:HDIconsistency}
    \nu_{HDI,V}^{INC}(t) = \| INC^{HDI,V}(t) \|; \\
    \nu_{GDP,V}^{INC}(t) = \| INC^{GDP,V}(t) \|.
    \label{eq:GDPconsistency}
\end{align}

\begin{figure}
    \centering
    \includegraphics[width=\textwidth]{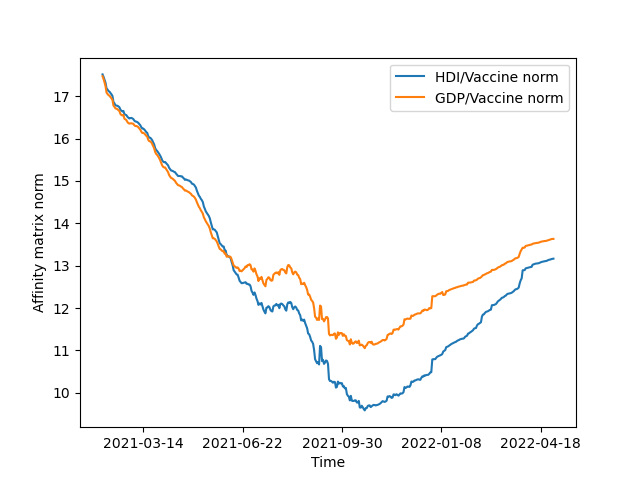}
    \caption{Time-varying collective consistency between vaccine proliferation and economic/human development, as defined in (\ref{eq:HDIconsistency}) and (\ref{eq:GDPconsistency}). There are several points of interest in the graph. First, both functions share a very similar evolution. Second, we see a global minimum at near-identical points, reflecting the similar {consistency} between vaccine proliferation and either HDI or GDP. Finally, the fact that HDI/vaccine consistency is regularly lower than GDP/vaccine consistency suggests that there is greater consistency between countries' human development index and the speed and progress of their vaccine rollout.}
    \label{fig:Time_varying_consistency}
\end{figure}

In Figure \ref{fig:Time_varying_consistency}, we display both $\nu_{HDI,V}^{INC}(t)$ and $\nu_{GDP,V}^{INC}(t)$ as a function of time.  These curves allow one to identify points in time where levels of economic and human development are most strongly  {consistent with} the proportion of the population that is fully vaccinated. This is marked by the time where both functions are minimal, namely where there is the least {collective} inconsistency (or greatest {collective} consistency) between economic/human development and country vaccine proliferation.

It is notable throughout our period of analysis that both {collective inconsistency} norm functions {(\ref{eq:HDIconsistency}) and (\ref{eq:GDPconsistency})} share a similar evolution, with both curves exhibiting a concave-up shape. That is, both  {comparisons} start with relatively high inconsistencies, decrease {in collective inconsistency} during the initial stage of our analysis, and {subsequently increase}. These two curves reveal several findings of interest. First, the broadly similar trajectories of the two curves indicate that GDP and HDI are broadly similar in their time-varying {collective consistency} with vaccine proliferation. They share the same down-up paths and global minima at very similar dates. Second, given that the HDI/vaccine inconsistency norm function reaches a lower point than the GDP/vaccine inconsistency norm, one could argue that HDI is more closely {consistent} with vaccine {proliferation} than GDP. Third, there is an interesting plausible explanation for the concave-up shape of each trajectory. Initially, when very few countries had administered their COVID-19 vaccination program, large discrepancies between GDP/HDI and vaccination rollout come from natural variability between countries' level of economic and human development. The point at which both inconsistency norm functions are lowest, which indicates the greatest level of consistency, {could} be explained by the more developed countries having mostly completed their vaccination programs. By the end of our window of analysis, when inconsistency matrix norms return close to their original values, less developed countries had begun to administer their COVID-19 vaccination program, and inconsistencies are driven by natural variation between countries' GDP and HDI values. {We expand on this possible interpretation in Section \ref{sec:Conclusion}.}

More broadly, this method of analysis could be used for exploring time-varying consistency between such attributes at a finer level. For example, one could subset two distinct collections of developed vs developing countries and analyse the extent of consistency between HDI/GDP and vaccine proliferation within each subsetted group.

\section{Discussion and conclusion}
\label{sec:Conclusion}

This paper has studied two COVID-19 multivariate time series on a country-by-country basis pertaining to data sets seldom studied in tandem (or even at all). Each section consists of mathematical approaches inspired from other fields such as financial correlations, graph theory and matrix algebra, with numerous insights attained.

\subsection{Paper summary}

In Section \ref{sec:Time_varying_infectiousness}, we explore the time-varying collective infectivity of the COVID-19 pandemic, from its inception until present day. To do so, we implement time-varying principal components analysis and study the evolution of the strength of the correlation matrix's first eigenvalue. Our analysis identifies a period corresponding to early 2021, where we see an abrupt translation upward in the collective strength of infectivity dynamics among the countries we have studied. The second key finding is the identification of periodic (oscillatory) trends in collective infectivity dynamics. This periodic behaviour could be  associated with two key phenomena: the emergence of more infectious strains of COVID-19 (such as the Delta variant), and the ``stop-start'' nature of various countries' economic reopenings. We complement the correlation analysis with an investigation of the collective magnitude of $R_t$ values across the world. We observe an association between the onset of the omicron variant and collective magnitudes of reproduction numbers, but not with collective correlations. Essentially, we have revealed that a new saga of similarity between  reproduction number time series among countries predated the virulent variant and might have more to do with countries overwhelmingly ``giving up'' containment measures in 2021 and reducing restrictions in large numbers \citep{US_lift,UK_lift,Europe_lift,Sweden_lift,Israel_lift}.

In Section \ref{sec:Graph_theoretic_partition}, we introduce a novel graph-theoretic framework to investigate and quantify the location of the change point identified in Section \ref{sec:Time_varying_infectiousness}. Our graph theoretic framework confirms that this transition occurs in March/April 2021. We then introduce two matrices $d(s,t)$ and $d^{10}(s,t)$ that explore the evolutionary similarity in the eigenspectrum of the infectivity correlation matrix. Both matrices present findings that are broadly consistent, with the emergence of two noticeably concentrated sub-matrices which display limited affinity between the two.

In Section \ref{sec:Vaccine_rollout_consistency}, we study the time-varying collective consistency between countries' economic and human development (represented by HDI and GDP) and their COVID-19 vaccine rollouts. To do so, we construct a variety of temporally-dependent affinity matrices, and explore the evolution of collective consistency. Our analysis confirms that our collection of countries shares a similar evolution in consistency between HDI and GDP with their respective accumulation of fully vaccinated individuals. Both trajectories share a global minimum in collective inconsistency, that is a maximum in collective consistency, which is virtually identical. We observe a potential association between this maximum in collective consistency and a point in time where more developed countries had displayed greater efficiency in their vaccine rollout. Finally, one can observe that there is greater consistency throughout the period of analysis between countries' economic development (HDI) and their vaccine proliferation when compared to their economic development (GDP) and their vaccine proliferation. Future work could use this method of analysis for exploring time-varying consistency between such attributes at a finer level, for example considering two distinct collections of developed and developing countries. In the same vein, our work could be used on a more local level, such as studying COVID-19 at the level of cities and localities, a growing field \cite{hancean_role_2021,Arajo2023,Sunahara2023_Matjaz}.

\subsection{Strengths and limitations}

In this paper, we have analysed the multivariate time series of reproduction numbers among a collection of 50 countries. Such reproduction numbers are ultimately a function of the confirmed COVID-19 case counts in each country, and thus require caution and discussion of limitations. Not only were many cases under-reported, particularly early on \citep{underreporting}, but testing protocols have been far from uniform over time and between countries. Indeed, several countries have changed their testing protocols on various occasions, including within the same wave of the outbreak \citep{Francechange,Pullano2020,Antigenchange}. Within Italy, for example, different regions within the same country operated according to different protocols, testing only symptomatic patients or more broadly \citep{DiBari2020}. Among our collection of 50 countries, differences in testing protocols, testing availability and even case definitions only widen \citep{Mercer2021}. To compare COVID-19 outcomes between two countries (or even two regions) in a traditional statistical model would require care to adjust for the difference in case reporting. In addition, (case) reproduction numbers are necessarily time-delayed estimates of the true instantaneous reproduction numbers, on the order of 1-2 weeks at most \citep{Pasetto2021}.

Fortunately, these limitations are not insurmountable to fruitful analysis, particularly in our setting, for several reasons. Principally, we do not analyse case counts directly, but effective reproduction numbers $R_t$ as estimated by \cite{arroyo-marioli_tracking_2021}. In Appendix A.6, these authors carefully discuss their methodology's sensitivity to reporting issues, and conclude that their methodology is relatively accurate to a range of potential data issues. They reason both via theoretical argument and Monte Carlo simulation. For example, if there is a constant detection rate $\alpha$, their calculation of $R_t$ would be unbiased. In the event of constant growth in the detection rate, their estimate is biased upwards or downwards if the testing growth rate is positive or negative, respectively; however, they argue that the trend in $R_t$ is of greater interest and is still estimated correctly; ``Intuitively, constant growth in the detection rate leads to a level bias, but the slope is still estimated correctly.'' The same applies in our analysis: as we calculate correlations between reproduction number time series over 90 day periods, trends matter more than raw values. As we use 90-day periods, we must acknowledge that our findings only have the precision of a monthly scale, such as the transition point in March/April 2021 determined by Sections \ref{sec:Time_varying_infectiousness} and \ref{sec:Graph_theoretic_partition}. However, this is no problem, and indeed inevitable given the aforementioned delays in case reproduction numbers.

In fact, the same reasoning would apply to raw confirmed case counts even without the calculations performed by \cite{arroyo-marioli_tracking_2021} to estimate the reproduction number. When calculating correlations, the trends matter more than the individual values, and one can gain inference even if some individual values are off, as long as there are noticeable trends in case counts. And indeed, case counts rarely resemble random sampling from distributions, but have clear trends, usually rising or falling in waves \citep{james2020covidusa,jamescovideu}. Perhaps most importantly of all, when computing collective strength via matrix norms, $\tilde{\lambda}_1(t)$ in (\ref{eq:lambda1}) or  $\nu_R(t)$ in (\ref{eq:R0vector}), individual discrepancies matter even less. Throughout this paper, we only ever analyse such collective measures across 50 countries.

In studying the collective behaviours among reproduction numbers, we must also acknowledge that the global state of COVID-19 evolution and prevalence may not necessarily be homogeneous. Over time, several variants have emerged, and spread around the world from their points of origin \citep{HadjHassine2021}. While they may affect their country of origin initially more than anywhere else, they quickly spread around the world \citep{Zhao2022_variants}. Thus, the introduction of new variants does not negate the utility to studying the \emph{collective} strength of reproduction numbers around the world. A new variant spreading in just one country will impact these collective strengths minimally; it will impact the collective strength of either correlations or absolute magnitudes more as it dominates a large number of countries. Expressed alternatively, nowhere in the paper do we directly compare and contrast two countries' COVID-19 epidemiology at any time in the period of analysis - instead, we always examine collective behaviours across 50 countries. Nor do we ever use a traditional statistical model to measure differences in outcomes between any two countries, so carefully equalising the COVID-19 situations between two countries is not necessary - we only examine collective trends across all countries. This is inspired among other things by the study of collective behaviours in financial markets, where profound idiosyncrasy may exist between different stocks, but substantial inference is possible when examining collective strength of correlations and other mathematical quantities \citep{Fenn2011,Heckens2020}.

That is, the emergence of a new COVID-19 variant is just one of many contributing factors towards the level of infectivity in a country at any time. Other contributing factors could be the level of government restrictions, among many others. Our paper is mathematical in nature, investigating time-varying collective dynamics with novel mathematical approaches, rather than a statistical model intending to isolate specific explanations. We never aim to explain specifically the causes for changes in trends - that is an opportunity for future research conducted with far more data on the local level. Thus, the emergence of different variants may be among the many reasons influencing the trends in infectivity, but these reasons are not the focus of this paper, instead the original mathematical analysis of collective trends.

The other multivariate time series of interest was the proportion of fully vaccinated individuals. While these data are likely to be more accurate than the confirmed case counts (attempting to track the true infection numbers), they are also worthy of comment. In Section \ref{sec:Vaccine_rollout_consistency}, we analyse the time-varying collective infectivity between countries' vaccine rollout and their HDI and GDP. It is worth commenting on such a relationship in broad terms. Some relationship is expected, because wealthier and more developed countries were able to secure supplies of COVID-19 vaccination before their less wealthy counterparts \citep{NYT_vaccine}. However, by no means is this a guaranteed or direct relationship, as political decisions, localised idiosyncratic vaccine hesitancy, or media reporting had a role in the ability or choice of the population to get vaccinated \citep{buonomo_behavioural_2022_jtb}. However, this is not necessarily detrimental to our analysis. Inconsistency between vaccine rollout and GDP or HDI is therefore expected to exist, and is worth analysing in a time-varying capacity. We observe an approximate concurrence in time between the period of the lowest collective inconsistency between countries' vaccine rollouts and human or economic development and a a time when developed countries have mostly completed their rollouts but less developed countries have mostly not. Nonetheless, alternative explanations or interpretations could exist and further examination is necessary.

We are unaware of any work that examines collective trends in reproduction number and vaccine proliferation throughout the pandemic using techniques from a broad range of areas including nonlinear dynamics and graph theory. In this work, we apply such methods to study a range in collective trends throughout the pandemic, including collective strength of correlations and magnitudes between reproduction number time series and collective consistency between vaccine proliferation and human/economic development indicators. Further work that more closely explores the key times - points of notable change or local maxima/minima - that we have identified in these collective trends would be welcomed by the community. Such analysis could help us better understand epidemiological dynamics during pandemics, and allow policymakers to implement better, data-driven decisions for optimal outcomes in the future.

\section*{Data availability statement}
All data are publicly available at Our World in Data (\url{https://ourworldindata.org}).

\section*{Funding statement}
No specific funding was received for this manuscript.

\bibliography{__CURRENTrefs.bib}
\bibliographystyle{_elsarticle-num-names}
\biboptions{sort&compress}
\end{document}